\documentclass[times]{cpeauth}

\usepackage{amssymb}
\usepackage{amsmath}
\usepackage{amsfonts}
\usepackage{algorithmicx}
\usepackage{algorithm}
\usepackage{algpseudocode}

\usepackage{graphicx}
\usepackage{url}
\usepackage{mathpartir}
\usepackage{tabularx}
\usepackage{multirow}
\usepackage{subfigure}
\usepackage{listings}
\usepackage{xspace}
\usepackage[latin1]{inputenc}

\newcommand{\esc}{e-Science Central\xspace}

\newcommand{\Div}{D$_{i,v}$\xspace}

\newcommand{\ED}{\mathit{ED}}
\newcommand{\wfms}{\mathit{wfms}}
\newcommand{\exec}{\mathit{exec}}
\newcommand{\tr}{\mathit{tr}}

\newcommand{\USED}{\mathit{used}}
\newcommand{\GENBY}{\mathit{genBy}}
\newcommand{\PDIFF}{{\sc PDIFF}}

\newcommand*\Call[2]{\textproc{#1}(#2)}

\title{Provenance and data differencing for workflow reproducibility analysis}

\author{Paolo  Missier\affil{1}\corrauth\ , Simon Woodman\affil{1}, Hugo Hiden\affil{1}, Paul Watson\affil{1}}
\address{\affilnum{1} School of Computing Science, Newcastle University, Newcastle-Upon-Tyne, UK}

\corraddr{E-mail: Paolo.Missier@ncl.ac.uk}

\begin{abstract}

One of the foundations of science is that researchers must publish the methodology used to achieve their results so that others can attempt to reproduce them. This has the added benefit of allowing methods to be adopted and adapted for other purposes. In the field of e-Science, services -- often choreographed through workflow, process data to generate results. The reproduction of results is often not straightforward as the computational objects may not be made available or may have been updated since the results were generated. For example, services are often updated to fix bugs or improve algorithms. This paper addresses these problems in three ways. Firstly, it introduces a new framework to clarify the range of meanings of ``reproducibility". Secondly, it describes a new algorithm, \PDIFF, that uses a comparison of workflow provenance traces to determine whether an experiment has been reproduced; the main innovation is that if this is not the case then the specific point(s) of divergence are identified through graph analysis, assisting any researcher wishing to understand those differences. One key feature is support for user-defined, semantic data comparison operators. Finally, the paper describes an implementation of \PDIFF~that leverages the power of the e-Science Central platform which enacts workflows in the cloud. As well as automatically generating a provenance trace for consumption by \PDIFF, the platform supports the storage and re-use of old versions of workflows, data and services; the paper shows how this can be powerfully exploited in order to achieve reproduction and re-use.

\end{abstract}

\begin{document}

\runningheads{P. Missier, S. Woodman, H, Hiden, P. Watson}{Provenance and data differencing for reproducibility}


\maketitle

\section{Introduction}  \label{sec:intro}

\subsection{Motivation}

A key underpinning of scientific discourse is the ability independently to verify or refute experimental results that are presented in support of a claim.
With e-science in particular, such a need translates into expectations that the computational machinery used by one research group to produce the results be made publicly available in sufficient detail as to be used by third parties to construct a scientific argument in support of, or against, those results.
Whilst it has long been the practice of scientific scholarly communication to require experimental methods to be described alongside the experimental data, it is generally accepted that reproducing an experiment based on its paper description requires substantial effort from independent researchers, either in the lab or in form of new program implementation.

Expectations regarding reproducibility in science have, however, been evolving towards higher levels of automation of the validation and reproduction process. This is driven by the observation that actual acceleration of the production of scientific results, enabled by technology infrastructure which provides  elastic computational facilities for ``big data'' e-science, is critically dependent upon the large-scale availability of datasets themselves, their sharing, and their use in collaborative settings. 

Following this observation, new social incentives are being tested with the goal of encouraging, and in some cases forcing, the publication of datasets associated with the results published in papers. Evidence of increasing sharing can be found mostly in the form of online repositories. These include, amongst others, the recently created Thomson's Data Citation Index\footnote{\url{wokinfo.com/products_tools/multidisciplinary/dci/}}; the Dryad repository for published data underlying articles in the biosciences (\url{datadryad.org/}); the databib.org registry of data repositories (\url{databib.org/}) (over 200 user contributed repository entries at time of writing), as well as 
Best Practice guidelines for data publication, issued by data preservation projects like DataONE\footnote{\url{notebooks.dataone.org/bestpractices/}}.

This dissemination phenomenon combines with the increasing rate at which publishable data sets are being produced, resulting in the need for automated validation and quality control of the experimental methods and of their outcomes, all of which imposes new reproducibility requirements. Here the term \textit{reproducibility} denotes the ability for a third party who has access to the description of the original experiment and its results to reproduce those results using a possibly different setting, with the goal to to confirm or dispute the original experimenter's claims.
Note that this subsumes, and is hopefully more interesting, than the simpler case of \textit{repeatability}, i.e. the attempt to replicate an experiment and obtain the same exact results when no changes occur anywhere.

In this paper we focus on the specific setting of workflow-based experiments, where reproduction involves creating a new version of the original experiment, while possibly altering some of the conditions (technical environment, scientific assumptions, input datasets). In this setting, we address the need to support experimenters in comparing results that may diverge because of those changes, and to help them diagnose the causes of such divergence.
Depending on the technical infrastructure used to carry out the original experiment, different technical hurdles may stand in the way of repeatability and reproducibility. The apparently simple task of repeating a completely automated process that was implemented for example using workflow technology, after a period of time since the experiment was originally carried out, may be complicated by the evolution of the technical environment that supports the implementation, as well as by changes in the state of surrounding databases, and other factors that are not within the control of the experimenter. This well-known phenomenon, known as ``workflow decay'', is one indication that complete reproducibility may be too ambitious even in the most favourable of settings, and that weaker forms of partial reproducibility should be investigated, as suggested by De Roure \textit{et al.}~\cite{Roure2011}. Yet, it is easy to see that this simple form of repeatability underpins the ability to reuse experiments designed by others as part of new experiments. Cohen-Boulakia \textit{et al.} elaborate on reusability specifically for workflow-based programming~\cite{Cohen-Boulakia:2011:SAR:2034863.2034865}: a older workflow that has ceased to function is not a candidate for reuse as part of a new workflow that is being designed today.

Additional difficulties may be encountered when more computational infrastructure with specific dependencies (libraries, system configuration) is deployed to support data processing and analysis. For some of these, well-known techniques involving the use of virtual machines (VMs) can be used. VMs make complex applications self-contained by providing a runtime environment that fulfills all of the applications dependencies. One can, for instance, create a VM containing an application consisting of a variety of bespoke scripts, along with all the libraries, databases and other dependencies that ensure the scripts can run successfully regardless of where the VMs are deployed, including on public cloud nodes.

\subsection{Goals}  \label{sec:goals}

VM technology is useful to provide a self-contained runtime environment, and indeed Groth \textit{et al.} show that a combination of VM technology for partial workflow re-run along with provenance can be useful in certain cases to promote reproducibility~\cite{Groth2009f}. However, our contention is that it is \textit{the capability  to compare execution traces and datasets with each other} that sits at the core of any argumentation that involves the reproduction of scientific results. Simple as it sounds, this aspect of reproducibility seems to have received relatively little attention in the recent literature, with a few exceptions as noted later. Yet, without appropriate operational definitions of \textit{similarity of data}, one cannot hope to answer the question of whether the outcome of an experiment is a valid confirmation of a previous version of the same, or of a similarly designed experiment. The work described in this paper is focused primarily on this aspect of reproducibility, namely on adding data-comparison capability to an existing e-science infrastructure, and to make it available to user scientists in addition to other reproducibility features, as a tool for supporting scientific discourse.

Our long term goal is to support experimenters in answering two related questions. 
Firstly, given the ability to carry out (variations of) the same experiments, to what extent can the outcomes of those experiments be considered \textit{equivalent}? Secondly, if they are not equivalent, what are the possible causes for the divergence?

Answering the first question is hard in general, as it requires knowledge of the type of the data, their format, and more importantly, their semantics. Consider for example two workflows, which process data in a similar way but at some stage employ two different statistical learning techniques to induce a classification model. Presumably, their outputs will be different for any execution of each of the two, and yet they may still be sufficiently close to be comparable. Thus, in this case a legitimate question is, are there suitable measures of data similarity that help support high-level conclusions of the form ``the second method confirms the results of the first on a specific set of input datasets''? Answering this question requires knowledge of machine learning theory, namely to establish whether two classifiers can be considered equivalent, although the results they produce are not identical. 
While the general problem of establishing data equivalence is knowledge-intensive, limiting the scope to simpler structural comparison may still lead to  effective conclusions. For instance, for some applications that operate on XML data, it may be enough to just define measures of similarity between two XML documents based on computing their structural differences, a  problem successfully addressed by Wang \textit{et al.}~\cite{xdiff}.
These examples suggest a spectrum of complexity in data similarity, starting from the simplest (two datasets are the same if their files are byte-wise identical), to structurally more complex data structures (XML ``diff'' operators), to end with the harder problems of semantics-based similarity as in the earlier machine learning example.

Our second question, regarding the diagnosis of the possible causes for dissimilarity of results, requires the ability to capture and then analyse the provenance traces associated with each of the experiments. Stucturally, the provenance trace of a workflow execution is an acyclic digraph (DAG) in which the nodes denote either a piece of data or an activity, and the edges denote relationships amongst data and activities, which reflect the production and consumption of data throughout the workflow execution steps (a simple but more formal definition is given in Sec.~\ref{sec:framework}). While the problem of comparing two traces is in principle a classic one of finding subgraph isomorphisms in directed labelled graphs~\cite{Berztiss:1973:BPI:321765.321766,Ullmann:1976:ASI:321921.321925}, greatly simplifying assumptions can be made by observing that workflow provenance traces are a specific type of acyclic digraph with additional labelling, namely regarding the ports on which activities produce and consume data items, which make it possible to reduce the space of potential matches.
Our assumption is that two provenance traces pertaining to two similar workflows are expected to be structurally similar, and to contain data nodes that are comparable, in the sense clarified above. When this is the case, then there is hope that one can detect points in the traces where results begin to diverge, and to diagnose the causes of such divergence by pairwise graph traversal.

\subsection{Contributions}

We offer two main contributions, both set in the context of workflow-based e-science. Our reference implementation is based upon the e-Science Central workflow management system, developed by Hiden \textit{et al.} in our group~\cite{Hiden2011}.
The first contribution (Sec.~\ref{sec:framework}) is a framework to characterise the \textit{reproducibility space}, defined by the change in time of one or more of the technical elements that make up a computational experiment, leading to various forms of ``workflow decay'', i.e. disfunctional or broken workflows.

The second contribution, described in Sec.~\ref{sec:provdiff-ddiff}, is \PDIFF, an algorithm and prototype implementation to help users analyse diverging outputs that occur when trying to reproduce workflows over time. This is complemented by an initial set of data comparison functions that address typical data types used in \esc workflows. As these workflows are routinely used to produce, for example, machine learning models for chemical engineering research~\cite{Cala2012}, our suite of data comparison functions includes the statistical similarity of predictive models produced using machine learning techniques (in addition to comparing plain text files and XML documents). This set of functions is easily extensible, as they are implemented using workflow blocks, which can be rapidly implemented and deployed on the middleware.

Finally, in Sec.~\ref{sec:infrastructure} we describe how \PDIFF~and the data diff functions leverage the current \esc middleware and complement the functionality of its existing provenance management sub-system. The technique described in the paper is not specific to the \esc platform, however. Indeed, since \esc provenance is encoded using the Open Provenance Model~\cite{Moreau2010a} (and will in the near future be aligned with the new PROV model~\cite{w3c-prov-dm}), it applies to any provenance system that adopts the standard.

\subsection{Related work}  \label{sec:related}

The perceived importance of enabling transparent, reproducible science has been growing over the past few years.
There is a misplaced belief, however, that adopting e-science infrastructures  for computational science will automatically provide longevity to the programs that encode the experimental methods. Evidence to the contrary, along with some analysis of the causes of non-reproducibility, is beginning to emerge, mainly in the form of editorials~ \cite{Hanson2011}, 
\cite{Merall2010}.  
The argument in support of reproducible research, however, is not new, dating back from 2006~\cite{Peng1-May-2006}. 
More recently, Dummond and Peng separately make a distinction between repeatability and reproducibility~\cite{Drummond2009}, \cite{Peng2011}. Bechhofer \textit{et al.} make a similar distinction concerning specifically the sharing of scientific data~\cite{Bechhofer2010a}.

In high profile Data Management conferences like SIGMOD, established initiatives to encourage contributors to make their experiments reproducible and shareable have been ongoing for a number of years, with some success\footnote{The SIGMOD reproducibility initiative: \url{www.sigmod.org/2012/reproducibility.shtml}.}.

Technical solutions for ensuring reproducibility in computational sciences date back over 10 years~\cite{Schwab2000}, but have matured more recently with workflow management systems (WFMS) like Galaxy~\cite{MesirovJill2010},  
\cite{Nekrutenko2010} 
and VisTrails~\cite{Scheidegger2008c}. In both Galaxy and VisTrails, repeatability is facilitated by ensuring that the evolution of the software components used in the workflows is controlled by the organizations that design the workflows. This ``controlled services'' approach is shared by other WFMS such as Kepler~\cite{Kepler1} and Knime (\url{www.knime.org/}), an open-source workflow-based integration platform for data analytics.
This is in contrast to WFMS like Taverna~\cite{Missier2010b}, which are able to orchestrate the invocation of operations over arbitrary web services, which makes it very general but also very vulnerable to version changes as well as maintenance issues of those services. A recent analysis of the decay problem, conducted by Zhao \textit{et al.} on about 100 Taverna workflows authored between 2007 and 2012 that are published on the myExperiment repository, reveals that more than 80\% of those workflows either fail to produce the expected results, or fail to run altogether~\cite{Zhao2012}. Such a high failure rate is not surprising given that experimental workflows by their nature tend to make use, at least in part, of experimental services which are themselves subject to frequent changes or may cease to be maintained. There are no obvious solutions to this problem, either, but the analysis technique described in this paper applies to Taverna provenance traces~\cite{Missier2010a}, which comply with the Open Provenance Model (and now with the W3C PROV model~\cite{w3c-prov-dm}).
 In this respect, \esc sits in the ``closed world'' camp, with versioned services data and workflows stored and managed by the platform. By default a workflow will make use of the latest version of a service, but this can be overridden, to the extent that past versions are still available on the platform. In any case, the version history of each service can be queried, a useful feature for the application described in this work.

The use of provenance to enable reproducibility has been explored in several settings. Moreau showed that a provenance trace expressed using the Open Provenance Model can be given a denotation semantics (that is, an interpretation) that effectively amounts to ``executing''  a provenance grpah and thus reproducing the computation it represents~\cite{Moreau2011202}. In the workflow setting, to the best of our knowledge, VisTrails and Pegasus/Wings~\cite{J.-Kim:2008kr} are the only WFMS where the evolution of workflows is recorded, leading to a complete provenance trace that accounts for all edit operations performed by a designer to create a new version of an existing workflow, as shown by Koop \textit{et al.}~\cite{springerlink:10.1007/978-3-642-17819-1_2}. A similar feature is available in \esc, namely in the form of a record history of all versions of a workflow. Whilst edit scripts can potentially be constructed from such history, those are currently not explicit or in the form of a provenance trace. Thus, our divergence analysis is solely based upon the provenance of workflow executions, rather than the provenance of the workflow itself. Note that this makes the technique more widely applicable, i.e. to the majority of WFMS for which version history is not available.

Our technique falls under the general class of graph matching \cite{Bunke2000}. Specifically relevant is the heuristic polinomial algorithm for change detection in un-ordered trees,  by Chawathe and Garcia-Molina \cite{Chawathe:1997:MCD:253262.253266}. Although matching provenance traces can be viewed as a change detection problem, it is less general because additional constraints on the data nodes, derived from the semantics of provenance graphs, reduce the space of possible matching candidates for any given node.

The idea of comparing workflow and provenance traces was explored by Altintas \textit{et al.}~\cite{DBLP:conf/ipaw/AltintasBJ06} for the Kepler system, with the goal of exploiting a database of traces as a cache to be used in lieu of repeating potentially expensive workflow computation wheneve possible. This ``smart rerun'' is based on matching provenance trace fragments with an upcoming computation fragment in a workflow run, and replace the latter with the former when a match is found. This serves a different purpose from \PDIFF, however. Whilst in SRM success is declared when a match is found, in \PDIFF~the interesting points in the graphs are those where divergence, rather than matching nodes, are detected. 

Perhaps the most prominent research result on computing the difference of two provenance graphs is that of Bao \textit{et al.}~\cite{Zhuowei-Bao:2009hq}. The premise of the work is that general workflow specifications that include arbitrary loops lead to provenance traces for which ``differencing'' is as complex as general subgraph isomorphism. The main insight is that the differencing problem becomes computationally feasible under the additional assumption that the workflow graphs have a series-parallel (s-p) structure, with well-nested forking and looping. While the \esc workflow model does allow the specification of non-s-p workflows, its loop facility is highly restricted, i.e. to a map operator that iterates over a list. Additionally, \PDIFF~relies on named input and output ports to steer the coordinated traversal of two provenance graphs, as explained in Sec.~\ref{sec:provdiff}. The combination of these features leads to a provenance comparison algorithm where each node is visited once.

%


\section{A framework for reproducibility in computational experiments}  \label{sec:framework}

The framework presented in this section applies to generic computer programs for which provenance traces can be generated. However, we restrict our attention to scientific workflows and workflow management systems, as the approach described in the rest of the paper has been implemented specifically for a workflow-based programming environment, namely eScience Central. Scientific workflows and their virtues as a programming paradigm for the rapid prototyping of science applications have been described at length by Deelman \textit{et al.}~\cite{Deelman2009528}. For our purposes, we adopt a minimal definition of workflow as a directed graph $W=\langle T, E \rangle$ consisting of a set $T$ of computational units (\textit{tasks}) which are capable of producing and consuming data items on \textit{ports} $P$ and a set $E \subset T \times T$ of edges representing data dependencies amongst processors, such that $\langle t_i.p_A, t_j.p_B \rangle \in E$ denotes that data produced by $t_i$ on port $p_A \in P$ is to be routed to port $p_B \in P$ of task $t_j$ during the course of the computation\footnote{More complete formalizations of workflow models are available, but this level of abstraction will suffice for our purposes.}.
We write:
\begin{equation}
\tr = \exec(W, d, \ED, \wfms) 
\label{eq:exec}
\end{equation}
to denote the execution of a workflow $W$ on input data $d$, with the additional indication that $W$ depends on external elements such as third-party services and database state, collectively denoted $\ED$ (for ``External Dependencies''). $W$ also depends on a runtime environment $\wfms$, namely the workflow management system.

The outcome of the execution is a graph-structured \textit{execution trace} $\tr$.
A trace is defined by a set $D$ of \textit{data items}, a set $A$ of \textit{activities}, and a set 
${\cal R} = \{R_1 \dots R_n\}$ of relations $R_i \subset (D \cup A) \times (D \cup A)$.
Different provenance models, including the OPM and PROV, introduce different types of relations. For the purpose of this work, we adopt a simple model ${\cal R} = \{ \USED, \GENBY \}$ where $\USED \subset A \times P \times D$ and  $\GENBY \subset D \times A \times P$ denote that an activity $a$ has used data item $d$ from its port $p$, written $\USED(a,p,d)$, and that $d$ has been generated by $a$ on port $p$, written $\GENBY(d,a,p)$, respectively.
Thus, a trace is simply a set of $\USED$ and $\GENBY$ records which are created during one workflow run, from the observation of the inputs and outputs of each workflow block that is activated. In particular, activities that appear in a trace correspond to instances, or invocations, of tasks that appear in the workflow.
Distinguished data items in a trace for a workflow run are the inputs and the outputs of the workflow, denoted $\tr.I$ and $\tr.O$ respectively. Formally, $\tr.I = \{ d \in D | a \in A \Rightarrow (d,a) \notin \GENBY$\}. Symmetrically, 
$\tr.O = \{ d \in D | a \in A \Rightarrow (a,d) \notin \USED$\}.

\subsection{Evolution of workflow execution elements}

At various points in time, each of the elements that participate in an execution may evolve, either independently or in lock-step. 
This is represented by subscripting each of the elements in (\ref{eq:exec}), as follows:
\begin{equation}
\tr_t = \exec_t(W_i, \ED_j, d_h, \wfms_k), \mbox{ with } i,j,h,k < t 
\label{eq:exec-evolve}
\end{equation}

\begin{figure}
 \centering
 \begin{center}
 \includegraphics[scale=.5]{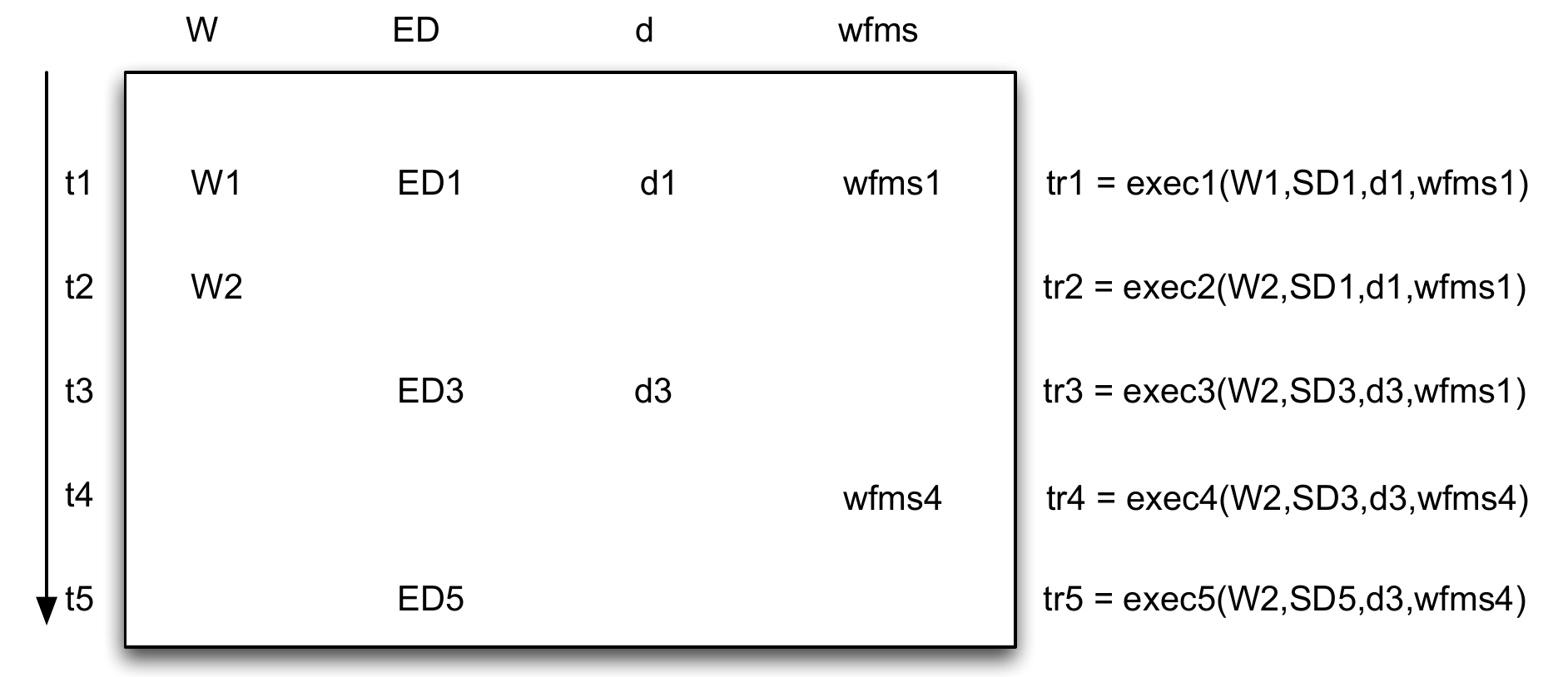}
\end{center}
 \caption{Example of evolution timeline}
 \label{fig:timeline}
\end{figure}

This indicates that a specific version of each of the elements that participate in the execution is used at time $t$.
Suppose for instance that version $1$ is initially defined at time $t_1$ for all elements:  $\langle W_1, \ED_1, d_1, \wfms_1 \rangle$. One execution at this time produces $\tr_1$. Then, at time $t_2$, a new version $W_2$ of the workflow is produced, everything else being unchanged. When executed, the workflow produces $\tr_2$.  At time $t_3$, further changes occur to the external dependencies and to the input dataset, leading to a new combination $\langle W_2, \ED_3, d_3, \wfms_1 \rangle$ and to the new result $\tr_3$. Fig.~\ref{fig:timeline} shows this incremental evolution, with changes happening at $t_4$ to the WFMS itself, from $\wfms_1$ to $\wfms_4$, and finally once more to the dependencies, from $\ED_3$ to $\ED_5$.
The last column in the figure indicates ``default'' execution, that is, executions based on the latest available version of each element.

We use this simple model to give a precise meaning to the informal notions of repeatability and reproducibility introduced earlier, as follows.
\begin{itemize}
\item 
We say that a baseline execution of the form (\ref{eq:exec-evolve}) is \textit{repeatable} if it is possible to compute (\ref{eq:exec-evolve}) at some time $t' > t$.
\item We say that execution  (\ref{eq:exec-evolve}) is \textit{reproducible} if it is possible to compute a new version of  (\ref{eq:exec-evolve}) at time $t' >t$, i.e.:
\[ \tr_{t'} = \exec_{t'}(W_{i'}, \ED_{j'}, d_{h'}, \wfms_{k'}) \]
  in the case where at least one of the elements  involved in the original execution at time $t$ has changed, i.e., $W_i \neq W_{i'}$, $ED_j \neq \ED_{j'}$, $d_h \neq d_{h'}$, or $\wfms_k \neq \wfms_{k'}$. 
  \end{itemize} 
  Notice that a situation where data changes: $d_h \rightarrow d_{h'}$ but $W_i$ remains the same, or vice versa, may require additional adapters to ensure that the new input can be used in the original workflow, or that the new workflow can accommodate the original input, respectively.

These simple definitions entail a number of issues that make repeating and reproducing executions complex in practice. Firstly, execution repetition becomes complicated when at least one the elements involved in the computation has changed between $t$ and $t'$, because in this case each of the original elements must have been saved and still made available.
Secondly, reproducibility is not trivial either, because in general there is no guarantee that the same workflow can be run with new external dependencies, or using a new version of the WFMS, or using a new input dataset.

Furthermore, we interpret the notion of reproducibility as a means by which experimenters validate results using deliberately altered conditions ($W_{i'}$, $d_{j'}$). In this case, the goal may not be to obtain exactly the original result, but rather one that is similar enough to it to conclude that the original method and its results were indeed valid. Thus, we propose that reproducibility requires the ability to \textit{compare the results $\tr_t.O$, $\tr_{t'}.O$ of two executions}, using dedicated similarity functions, i.e., $\Delta_D(\tr_t.O, \tr_{t'}.O)$ (for example, one such function may simply detect whether the two datasets are identical).
However, a further complication arises as the criteria used to compare two data items in general depends on the type of the data, requiring the definition of a whole collection of $\Delta_D$ functions, one for each of possibly many data types of interest. 

A final problem arises when one concludes, after having computed $\exec_{t}(\dots)$, $\exec_{t'}(\dots)$, and $\Delta_D(\tr_t.O, \tr_{t'}.O)$, that the results are in fact not similar enough to conclude that the execution has been successfully reproduced. In this case, one would like to identify the possible causes for such divergence.

We address these problems in the rest of the paper. In particular, after discussing general techniques for supporting repeatability and reproducibility, we focus on the last problem, namely the role of provenance traces in diagnosing divergence in non-reproducible results, and describe a prototype implementation of such a diagnostic tool built as part of the eScience Central WFMS, underpinned by simple but representative data similarity functions.

\subsection{Mapping the reproducibility  space} \label{sec:mappingspace}

\begin{figure}
 \centering
 \begin{center}
 \includegraphics[scale=.5]{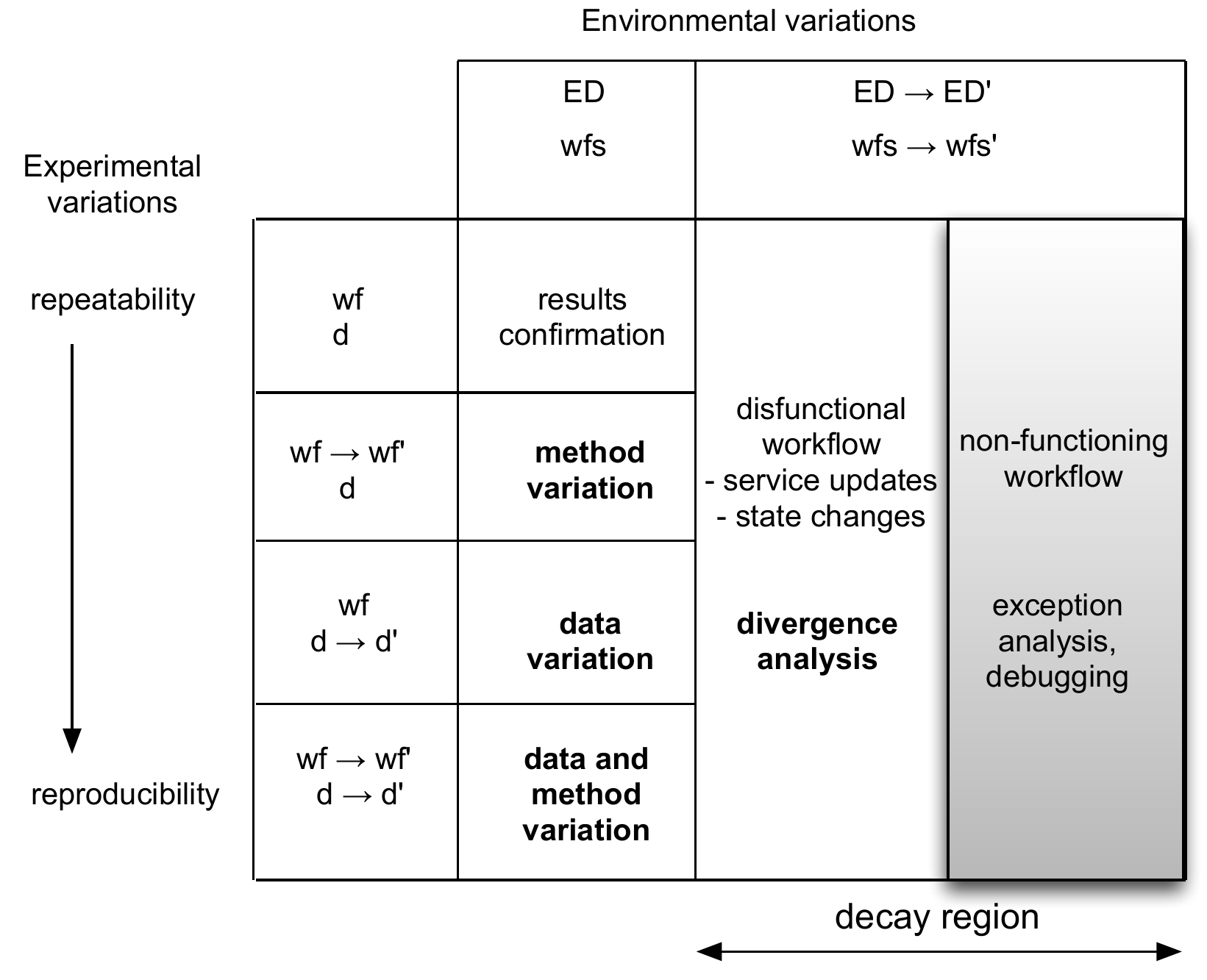}
\end{center}
 \caption{Mapping the reproducibility space}
 \label{fig:repro-space}
\end{figure}

Our formulation can be used to summarize some of the repeatability/reproducibility scenarios mentioned earlier. The grid in Fig.~\ref{fig:repro-space} illustrates a distinction between elements whose evolution is under the control of the experimenter, namely $W$ and $d$, and those that are beyond their control, namely $\ED$ and $\wfms$. Ignoring the latter for a moment, one can interpret the co-evolution of methods and data in terms of objectives that are part of the scientific discourse, ranging from simply repeating the experiment (as defined above), to observing the effect of changes in the method, in the input data, or both. Our implementation of a provenance-based divergence analysis with associated data similarity functions is aimed mainly at supporting the latter three scenarios.

When external factors evolve as well, various forms of \textit{decay} occur, leading to workflows that no longer deliver the intended functionality, or cannot be executed altogether. Our analysis tool may help diagnose the former case, where the results are not as expected and the cause is to be found in new versions of third-party services, updated database states, or the non-deterministic behaviour of some of the workflow blocks. In contrast, exceptions analysis is required to diagnose a broken workflow, and it is out of the scope of this work.
 
\section{Provenance and data differencing}  \label{sec:provdiff-ddiff}

\subsection{Detecting divergence in execution traces}  \label{sec:provdiff}

In this section we present \PDIFF, an algorithm designed to compare two provenance traces, $\tr_t$, $\tr_{t'}$, that are generated from related executions, i.e., of the form (\ref{eq:exec-evolve}) where at least one of the four contributing elements has changed between times $t$ and $t'$. Specifically, the algorithm attempts to identify differences in the two traces, which are due to one of the causes portrayed in Fig.~\ref{fig:repro-space}, namely workflow evolution, input data evolution, as well as non-destructive external changes such as service and database state evolution. Note that we only consider cases where both executions succeed, although they may produce different results. We do not consider the cases of workflow decay that lead to nonfunctional workflows, as those are best analysed using standard debugging techniques, and by looking at runtime exceptions.

\begin{figure}[htb]
 \centering
 \begin{center}
\includegraphics[scale=.4]{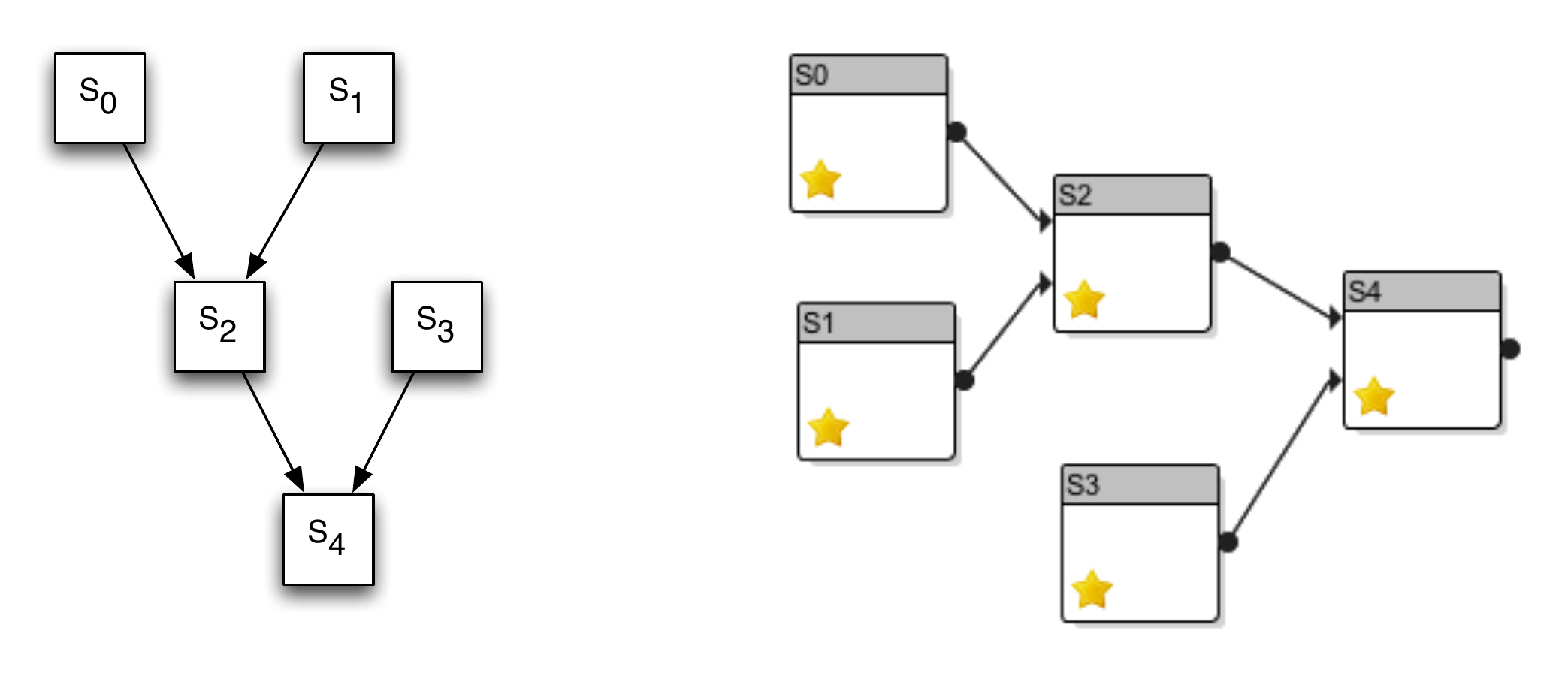}
 \end{center}
 \caption{Simple workflow (and its representation in the \esc GUI) to illustrate divergence analysis in the presence of data evolution}
 \label{fig:data-diff-workflow}
\end{figure}

\begin{figure}[htb]
 \centering
 \begin{center}
\includegraphics[scale=.4]{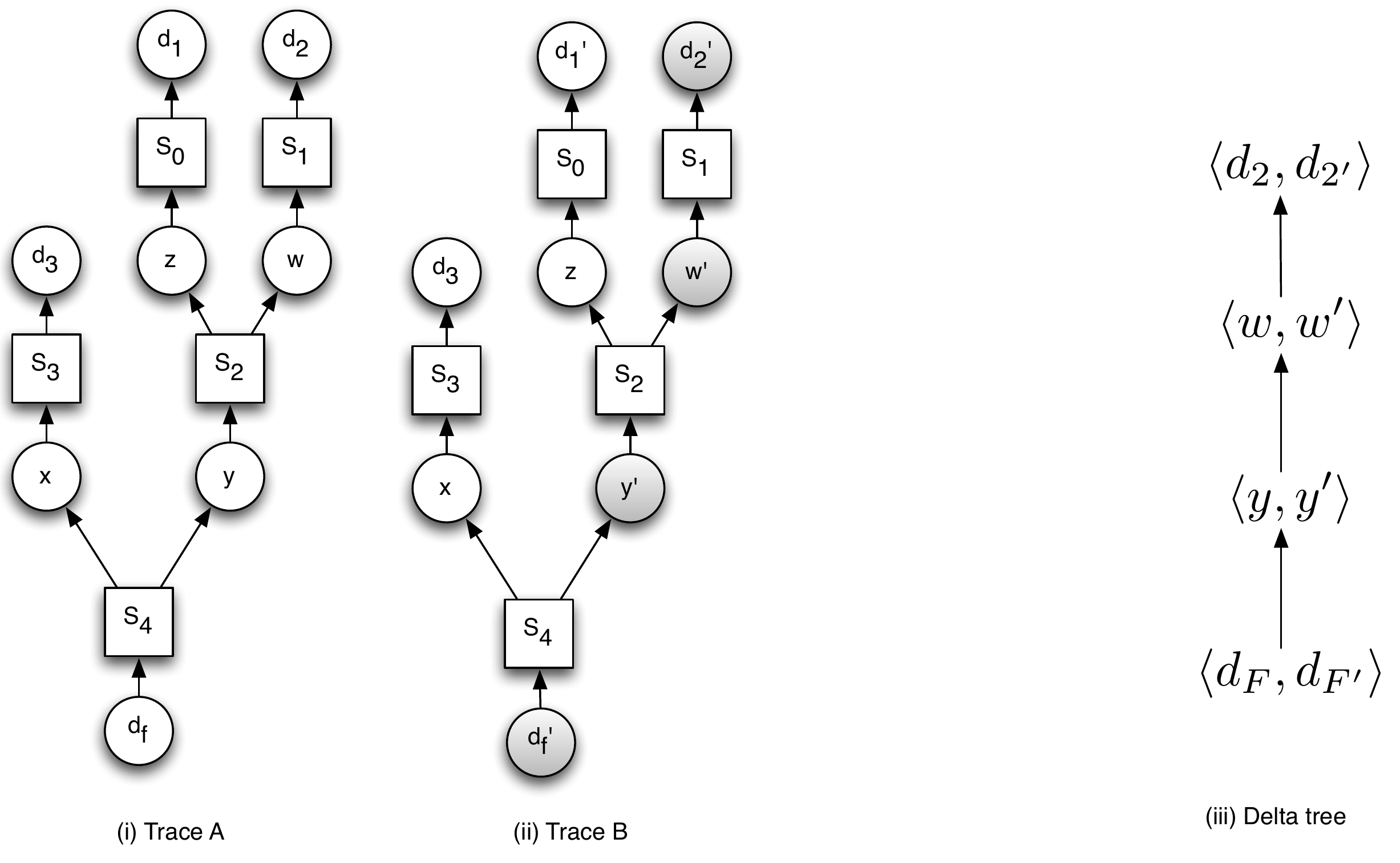}
 \end{center}
 \caption{Example traces for two executions of the workflow of Fig.~\ref{fig:data-diff-workflow}, and simple delta}
 \label{fig:data-diff-traces}
\end{figure}

The aim of the divergence detection algorithm is to identify whether two execution traces are identical, and if they are not, where do they diverge. Specifically, given two outputs, one from each of the two traces, the algorithm tries to match nodes from the two graphs by traversing them in lock-step as much as possible, from the bottom up, i.e., starting from the final workflow outputs. When mismatches are encountered, either because of differences in data content, or because of non-isomorphic graph structure, the algorithm records these mismatches in a ``delta'' data structure and then tries to continue by searching for new matching nodes further up in the traces. The delta structure is a graph whose nodes are labelled with pairs of trace nodes that are expected to match but do not, as well as with entire fragments of trace graphs that break the continuity of the lock-step traversal. Broadly speaking, the graph structure reflects the traversal order of the two graphs. A more detailed account of the construction can be found in Sec.~\ref{sec:support-functions}.
Our goal is to compute a final \textit{delta tree} that can be used to explain observed differences in workflow outputs, in terms of differences throughout the two executions.
We first illustrate the idea in the case of simple data evolution:
\[ \tr_t = \exec_t(W, \ED, d, \wfms),  \tr_{t'} = \exec_{t'}(W, \ED, d', \wfms) \]
The workflow for this example has three inputs and one output, and is portrayed in Fig.~\ref{fig:data-diff-workflow}. The two execution traces in Fig.~\ref{fig:data-diff-traces}, denoted A and B, correspond to inputs $tr_t.I = \{ d_1, d_2, d_3 \}$ and $tr_{t'}.I = \{ d_{1'}, d_{2'}, d_3 \}$, respectively. The two outputs $d_F, d_{F'}$ differ, as expected, however note that not all of the inputs contribute to this mismatch. Indeed, observe that input $d_{1'}$ on $S_0$ produces the same result as input $d_{1}$. Running \PDIFF~ on these two traces produces the delta graph, indeed a simple path in this case, shown in the rightmost part of Fig.~\ref{fig:data-diff-traces}. The path consists of pairs of mismatching nodes, that can be used to explain the mismatch $d_F \neq d_{F'}$. We can see for example that $d_2 \neq d_{2'}$ caused all of the intermediate data products downstream from $S_1$ to differ, eventually explaining the difference in the outputs.

\PDIFF~ works by traversing the two traces from the bottom up. As long as the two graphs are isomorphic, as is the case in this simple example, the algorithm proceeds in a breadth-first fashion up the traces, simply testing for differences in data and activity nodes, alternately. A new node is added to the delta graph whenever mismatches are encountered. The structure is a graph, rather than a tree, because different branches can sometimes be merged, as shown in the next example. Note that mismatches may occur even when the input datasets are identical, for example when one or more of the services exhibits non-deterministic behaviour, or depends on external state that has changed between executions.

\begin{figure}[ht]
 \centering
 \begin{center}
  \subfigure[] {
   \includegraphics[scale=.4]{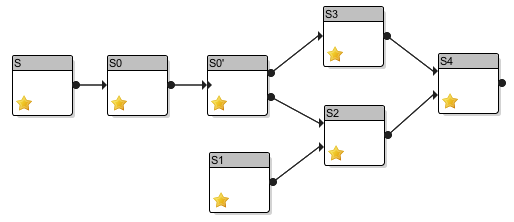}
   }   
     \quad
  \subfigure[] {
   \includegraphics[scale=.4]{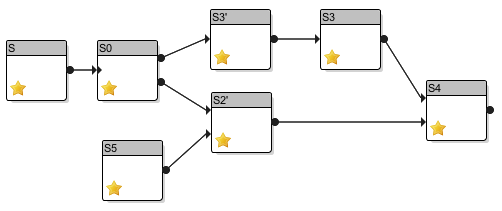}
   }   
 \end{center}
 \caption{Example workflows to illustrate general divergence analysis}
 \label{fig:any-diff-workflow}
\end{figure}

The case of non-isomorphic traces, which arise in the case of workflow evolution or of evolution in the version of some of the services, present additional challenges to the matching algorithm. As a second, more elaborate example, consider the two workflows in Fig.~\ref{fig:any-diff-workflow} (not that these are two different evolutions of the workflow of Fig.~\ref{fig:data-diff-workflow}). The main differences between them are as follows:
\begin{itemize}
\item $S_0$ is followed by $S_{0'}$ in $W_A$ but not in $W_B$;

\item $S_3$ is preceded by $S_{3'}$ in $W_B$ but not in $W_A$;

\item $S_2$ in $W_A$ is replaced by a new version, $S_{2v2}$, in $W_B$;

\item $S_1$ in $W_A$ is replaced by $S_{5}$ in $W_B$.
\end{itemize}

\begin{figure}[htb]
 \centering
 \begin{center}
 \includegraphics[scale=.4]{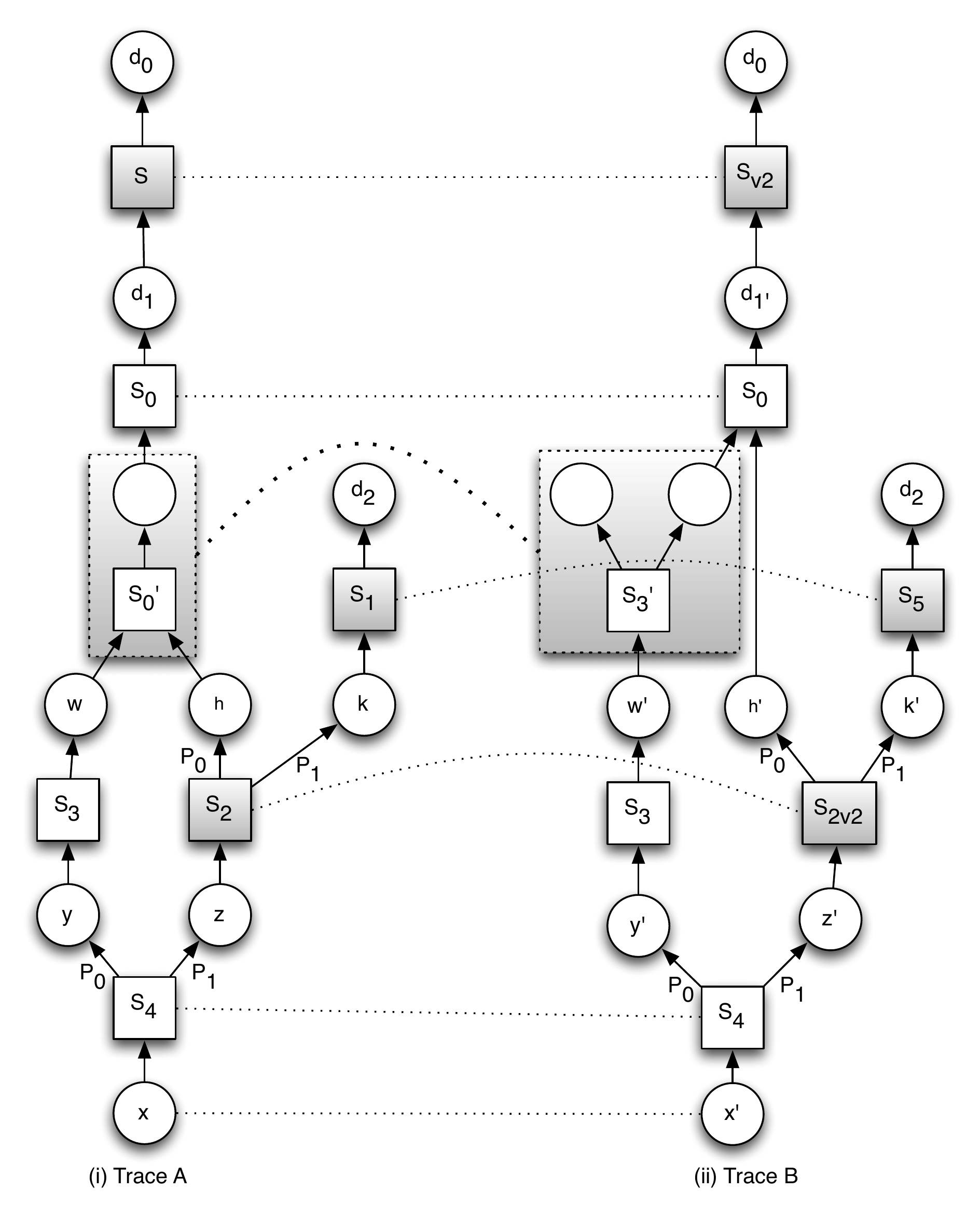}
 \end{center}
 \caption{Example traces for two executions of the workflow of Fig.~\ref{fig:any-diff-workflow}}
 \label{fig:any-diff-traces}
\end{figure}

\begin{figure}[htb]
 \centering
 \begin{center}
 \includegraphics[scale=.4]{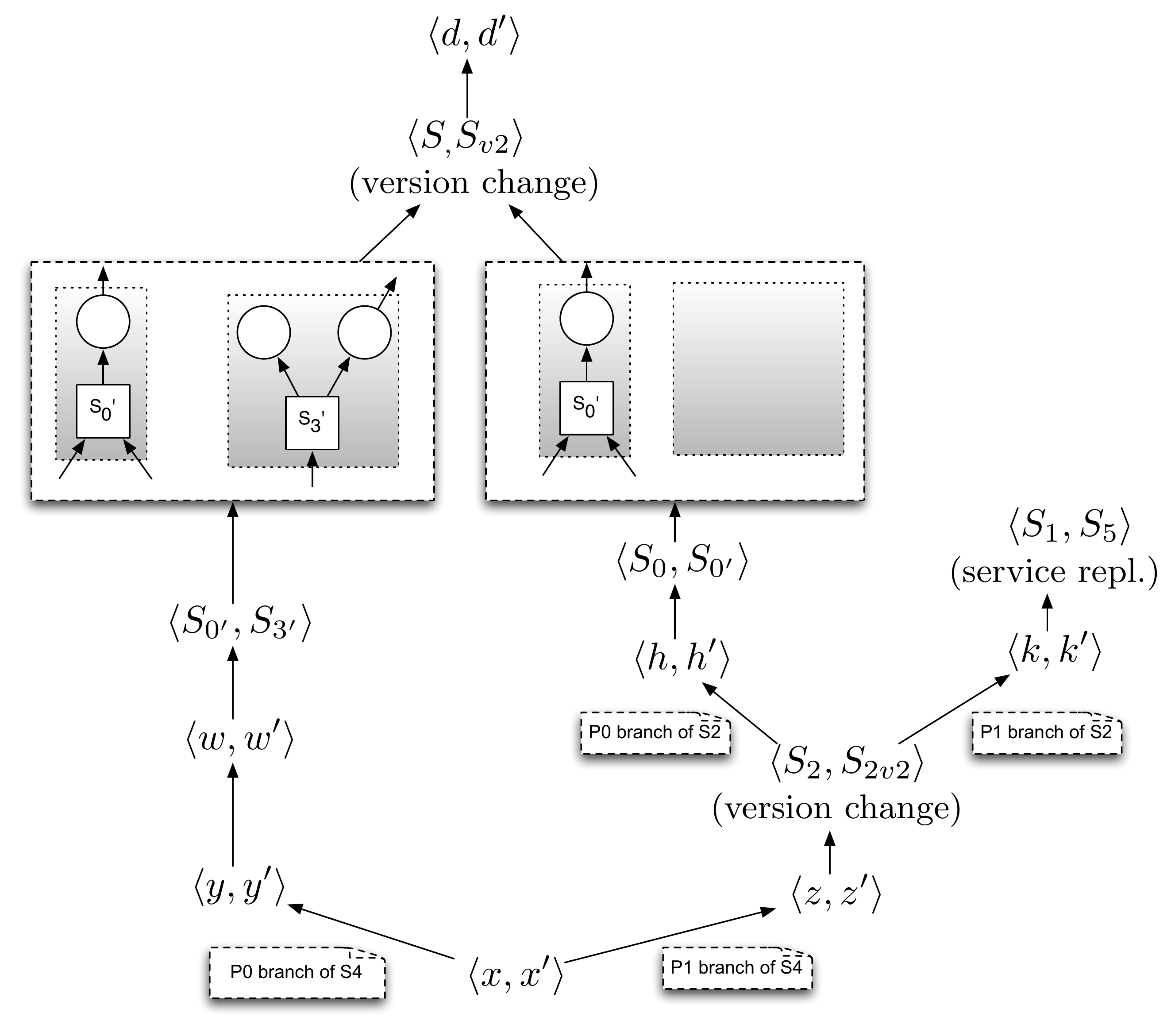}
 \end{center}
 \caption{Delta graph from \PDIFF~ on the traces of Fig.~\ref{fig:any-diff-traces}}
 \label{fig:any-diff-delta}
\end{figure}

Let us assume that the input data is the same for both workflows, resulting in the following scenario:
\[ \tr_t = \exec_t(W, \ED, d, \wfms),  \tr_{t'} = \exec_t(W', \ED', d, \wfms) \]
where $\ED'$ consists of the version change in service $S_2$.
\PDIFF~ operates on the assumption that the two traces are not too drastically different, and on the expectation that workflow evolution can be described using a combination of the patterns listed above. Although the algorithm provides no formal guarantees that all possible cases of graph structure mismatches will be correctly identified, this informal assumption is true for the two traces portrayed in Fig.~\ref{fig:any-diff-traces}, corresponding to one execution of each of the two example workflows. On these traces, \PDIFF~ produces the delta graph shown in Fig.~\ref{fig:any-diff-delta}. The algorithm operates as follows.

Initially, the outputs $tr_t.O = x$ and $tr_{t'}.O = x'$ are compared, and the initial mismatch $x \neq x'$ is recorded in the delta graph. 
Note that the outputs are uniquely determined by a combination of the unique workflow run, the ID of the output block that produces them, and the name of the output port.
Moving up, the services that produce $x$ and $x'$, $S_4$,  are found to be the same, and no action is taken. As $S_4$ has consumed two inputs\footnote{The algorithm assumes that the same service will always have the same number of ports.}, the traversal branches out to follow each of the two ports, $p_0, p_1$. Considering the left branch, the mismatching transient data $y \neq y'$ is detected and added to the delta graph. The traversal then continues past $w, w'$, when a service mismatch $S_{0'}, S_{3'}$ is detected. This is recorded in the delta graph, but it also triggers an exploration upwards on both traces, to find the nearest matching service node, if it exists, which would allow the pairwise comparison to resume. In this case, $S_0$ is such a ``sync'' node (it happens to be at the same level in both graphs in this example, but this is not necessarily the case). The graph fragments that were skipped during the search on each trace represent relevant structural differences in the two graphs, and are therefore recorded as part of the delta graph as well. These represent structural variations that occurred in each of the two graphs. In the next step, a version mismatch is detected between $S$ and $S_{v2}$, and again added to the graph. The final step on this branch finds identical original input data, $d_1$.

A similar traversal occurs on the right side of the traces, starting at port $p_1$ of $S_4$. The mismatch  $z \neq z'$ is first recorded, followed by $S_2 \neq S_{2v2}$. In this case, however, \PDIFF~is able to detect that the mismatch is due to a simple service version upgrade, as service versions are available as part of the provenance trace, and thus it does not trigger any further upwards search. One more branching occurs at this point: on the left, $h \neq h'$, followed by a further service mismatch, $S_{0'} \neq S_0$. As we saw earlier, this causes \PDIFF~to search for the nearest matching service node, which is again $S_0$. Note however that in this instance, the ``diff'' graph fragment on the right trace is null, because $S_{0'}$ has been added to one of the two workflows. The traversal resumes from $S_0$ on both sides. Because this node was used earlier as a sync point on the left branches of the two traces,  any further traversal upwards in the graph would follow exactly the same steps, and therefore it can be avoided. Furthermore, the delta path starting with pair $\langle S, S_{v2} \rangle$ can be shared with this branch of the delta graph, resulting in a join in the graph as shown at the top of Fig.~\ref{fig:any-diff-delta}.
 Finally, we come to the $p_1$ branch of $S_{2v2}$, where $k \neq k'$ and then $S_1 \neq S_5$. Although this does trigger a new upwards search, this ends immediately upon finding the input data.

We have mentioned that \PDIFF~assumes that the two traces are similar enough for this comparison to progress to the end. Large differences amongst the traces are manifested as divergence that cannot be reconciled later in the graph traversal. In this case, the earlier divergence point is reported. For instance, if the same output is produced by drastically different workflows, then \PDIFF~reports that there are no sync points at all.

\begin{algorithm}
\caption{\PDIFF: Provenance trace divergence detection}
\label{alg:pDiff}
\footnotesize
\begin{algorithmic}[0]

\Function{pDiff}{$\Delta, nL, nR$}

\If{$nL$ \textbf{is null or} $nR$ \textbf{is null}}
   \State \textbf{return} $\Delta$
 \EndIf
\If{$nL$.type == DATA\_NODE}     \Comment when comparing two data nodes:
  \If{ \Call{dDiff}{$nL,nR$}}   \Comment data mismatch
  	  \State $\Delta \gets $\Call{addDelta}{$\Delta, \langle nL, nR \rangle$}  \Comment add this mismatch to $\Delta$
  \EndIf
  \State \textbf{return} \Call{pDiff}{$\Delta, \Call{upD}{nL}, \Call{upD}{nR}$}  \Comment  and continue up the graph
\EndIf

\If{$nL$.type == ACTIVITY\_NODE}     \Comment when comparing two activity nodes:

  \If{ \Call{aDiff}{$nL,nR$}}   \Comment activity mismatch
     \If { (different version of the same service) }
         \State $\Delta \gets $\Call{addDelta}{$\Delta, \langle nL, nR \rangle$}  \Comment add this version mismatch to $\Delta$ and continue
     \Else
              \State $\Delta \gets $\Call{addDelta}{$\Delta, \langle nL, nR \rangle$}  \Comment add this service mismatch to $\Delta$
       \State $\langle GL, GR, nL', nR' \rangle \gets \mbox{\Call{findNode}{$nL,nR$}}$   \Comment resync traversal if possible
       \If{$nR'$ \textbf{is null} \textbf{and}  $nL'$ \textbf{is null}}    \Comment resync failed
            \State \textbf{return} $\Delta$ \Comment  return
       \EndIf
       \State $\Delta \gets $\Call{addDelta}{$\Delta, \langle GL, GR\rangle$}   \Comment resync succeeded, record mismatching graphs 
       \State $stopFlag \gets $\Call{isDeltaStop}{$\Delta$}   \Comment has a branch been joined in the delta graph?
       \If{$stopFlag$}  \State  \textbf{return} $\Delta$  \Comment sync'ed to a previous sync point, delta graphs joined. this traversal terminates
       \EndIf
       \If{ $nR$ \textbf{is not null}}        \State $nR \gets nR'$        \EndIf
       \If{ $nL$ \textbf{is not null}}        \State $nL \gets nL'$        \EndIf
   \EndIf
   \EndIf
   \State  \textbf{return}  \Call{MergeDeltas}{\{ \Call{pDiff}{$\Delta, \Call{upS}{nL,p}, \Call{upS}{nR,p}} | p \in nL.ports $ \} } \Comment traverse each input port
\EndIf  
\EndFunction
\newline

\end{algorithmic}
\end{algorithm}

\subsection{Support functions} \label{sec:support-functions}

The pseudo code for the algorithm appears in Alg.~\ref{alg:pDiff}, using a conceptual recursive formulation for clarity. The support functions used by the algorithm are described informally, as follows.

\begin{itemize}
\item Functions \textproc{upD}($d$) and \textproc{upS}($a,p$) perform one step of the upwards graph traversal. Specifically,  \textproc{upD}($d$) traverses a $\GENBY(d,a)$ relation where $d$ is a data node and returns the node for the activity $a$ that generated $d$.  \textproc{upS}($a,p$) traverses a $\USED(a,d,p)$ relation where $a$ is an activity node, and returns the node for the data $d$ used by $a$ through port $p$.

\item Function \textproc{dDiff}($nL, nR$) defines datatype-specific comparison of data. This is used when matching the inputs and outputs of two workflow executions, i.e., $\tr_t.I$ and $\tr_t.O$. Examples of such data comparison functions are presented in the next section. Note that intermediate results, represented as internal data nodes in the traces above, are transient data which are not explicitly stored in the provenance database. To save space, only their MD-5 hash is recorded in the node. This results in a crude but very efficient data mismatch test for internal nodes with boolean result. In contrast, in the case of actual data the function may produce a similarity value between 0 and 1. For simplicity, we assume that a threshold is applied, so that data diff functions all return {\sc true} if significant differences are detected, and {\sc false} otherwise\footnote{Whilst the transient nature of intermediate data does preclude the use of content-based comparison functions, note that one can control the granularity of provenance, and thus the points where actual data is stored with it, by breaking down a complex workflow into a composition of smaller ones. Using small workflows is indeed typical for existing \esc applications.}.
Similarly, function \textproc{aDiff}($nL, nR$) compares two service nodes using the service identifiers, and returns {\sc true} if the services are not the same. It can further flag differences that are due solely to mismatches in service versions.

\item \textproc{findNode}($nL, nR$) performs the search for a new matching node following a service node mismatch. It returns a 4-tuple consisting of the graph fragments that were skipped during the search (either or both can be null), as well as references to the matching service nodes in each of the two traces, if they exist (note that either both of these are valid references, or neither is). More formally,  
consider the subgraphs upwards from $nL$ and $nR$, i.e., $upL$ = \{\Call{upS}{$nL,p_i$}\}  for all input ports $p_i$ of $nL$, and 
    $upR$ = \{\Call{upS}{$nR,p_j$}\}  for all input ports $p_j$ of $nR$.
 \textproc{findNode} finds the closest matching activity nodes, i.e., the earliest encountered activity nodes 
    $nL'$ in $upL$ and $nR'$ in     $upR$
    such that  \Call{aDiff}{$nL',nR'$} is false.
    It records the mismatching nodes encountered in the new graph fragments $GL$, $GR$, respectively, or it returns null node identifiers if the search fails.

\item Finally, function \textproc{addDelta}($\Delta, deltaElement$) adds $deltaElement$ to  $\Delta$ and returns the new delta graph. This function also needs to be able to join tree branches on sync points as discussed in our previous examples. Suppose a delta node $g$ containing two graph trace fragments is added to the delta, and suppose $n$ is next sync node found by {\sc FindNode}. In this case, $g$ is annotated with $n$. If $n$ is found again at a later time following a new graph mismatch $g'$, then this is an indication that the traversal can stop, and that the graph above $g$ can be shared with $g'$. Fig.~\ref{fig:any-diff-delta} shows one example.  Auxiliary function \textproc{IsDeltaStop}($\Delta$) returns true if a merge has occurred, indicating that the current traversal may terminate.

 This function is generalized by \textproc{mergeDeltas}($deltaset$), which combines a set of $\Delta$ graphs into a new graph by making them siblings of a new common root. 

\end{itemize}

We now briefly analyse the complexity of \PDIFF. Let $N,M$ be the total number of nodes in the left and right graph, respectively.
The trace traversal is driven by functions \textproc{upD}() and \textproc{upS}(), which perform one step upwards in each of the two graphs. As noted, port names ensure that only matching $\GENBY()$ (resp. $\USED()$) relations are traversed. Thus, if the two traces were identical, \PDIFF~would complete the traversal in $N=M$ steps. On the other hand, a mismatch amongst two activity nodes may trigger a call to the \textproc{findNode}() operator, which involves searching for pairs of matching activity nodes across the two traces. Here port names cannot be used to assist the matching, which therefore requires one activity node lookup in one of the graph, for each activity node in the other graph. In the worst case, this results in $N . M$ comparisons for a call to \textproc{findNode}() that occurs near the bottom of the graph (i.e., at the first activity node from the bottom). As the traversal progresses upwards, more calls to \textproc{findNode}() may occur, but they operate on progressively smaller fragments of the graphs, and any new search will stop at the first sync node that was found in a previous search (i.e., when function \textproc{IsDeltaStop}() returns true). Considering the extreme cases, suppose that divergence is detected at the bottom of the graphs, and the only sync node is at the very top. This involves only one call to \textproc{findNode}(), with cost approximately $N . M$, and then \PDIFF~terminates. In this case, the algorithm involves at most $\mathit{min}(N,M) + N . M$ steps. The other extreme case occurs when the sync nodes are one step away from the diverging nodes. Now \textproc{findNode}() is called repeatedly, at most $\mathit{min}(N,M)$ times, but each time it completes after only one comparison, resulting in $2 . \mathit{min}(N,M)$ steps.

To summarize, \PDIFF~ is designed to assist reproducibility analysis by using provenance traces from different executions to identify the following possible causes of diverging output values for those executions:

\begin{itemize}
\item data and workflow evolution;

\item non-deterministic behaviour in some of the services, including the apparent non-determinism due to changes in external service (or database) state;

\item service version upgrades.
\end{itemize}

\subsection{Differencing data (DDiff)}  \label{sec:datadiff}

Determining the similarity between two data sets is highly dependant not only on the type and representation but also the semantics of the data.  Not all data sets can be compared using a byte-wise comparator to determine if they are equivalent.  For example, two models produced from the same data set may well be equivalent but are unlikely to be byte-wise identical. In addition, a Boolean response of equivalent or not equivalent is a somewhat na\"{i}ve approach and may provide false negatives and be less amenable to statistical analysis.

Within \esc we have built a flexible architecture in which it is possible to embed different algorithms for different types and representations of data.  Following the standard extensibility model of \esc, new data comparison functions can be implemented as workflow blocks (typically as Java classes that interact with the system's API).

We provide algorithms which calculate the equivalent of three classes of data.  The algorithms are implemented as \esc workflows (described in Section \ref{sec:infrastructure}) which implement a common interface allowing the system to  determine dynamically which one to use for the relevant data set.

\begin{description}
\item[Text or CSV]  Comparing textual files is straightforward, and tools such as GNU \texttt{diff} or \texttt{cmp}\footnote{gnu.org/software/diffutils/} are commonly used and well understood.  By default GNU \texttt{diff} works by comparing each file, line by line, and returning the differences, known as \emph{hunks}.  It is possible to configure \texttt{diff} to ignore whitespace or text case differences which is useful for scientific data sets where, for example, the header row may have different capitalisation but the data sets are otherwise identical.  If no \emph{hunks} are returned by diff the files are identical.  However, if \emph{hunks} are returned, the files differ in some way.  In order to calculate the similarity, $s$, we return the percentage of lines that are unchanged, converted to a number in the range $0 \leq s \leq 1$

\item[XML] Comparing XML files is more complex than comparing textual data due to the semantic rules governing its structure.  For example, the ordering of attributes is not important, generally the order of nodes is not important (although this is schema dependent) and where namespaces are used the prefix is not important.  Existing work on comparing XML documents often tries to minimise the transformation necessary to turn one document into another (the so called, `edit script') or maximise the common structure from one document to another \cite{conf/extreme/SchubertSB05,xdiff,xydiff}. For the purposes of comparing XML documents we use XOM (\url{xom.nu/}) which is able to canonicalise XML documents so that they can be compared.  Again, similarity, $s$ is the percentage of the document which has changed where $0 \leq s \leq 1$.

\item[Mathematical Models] Some modelling algorithms (for example, Neural Networks \cite{modelfit}) have an element of random initialisation and can be expected to give different results each time they are applied to the same dataset. Whilst these models may fail a byte-wise comparison, their predictive capabilities might be statistically the same. To calculate the similarity of models, we use the ANCOVA test (Analysis of Covariance - \cite{anova}) which  analyses the predictive performance of two models. This is demonstrated below in Figure \ref{fig:modelfit}, which shows a plot of two models' estimates of a variable against the actual observed values for that variable. A good model in this case would have a regression line at 45 degrees with very little spread in the actual and observed points. Different models applied to the same data set could be expected to have slightly altered regression line slopes, and the differencing tool needs to be able to take account of this. 

The ANCOVA test is divided into two parts: the first determines whether the slope of the regression lines is significantly different (A and B); whilst the second checks whether the y-axis intercepts of the two regression lines (C and D) are equivalent. It should be noted that whilst there are many more tests that should be done in order to compare two models (analysis of the model residuals, comparison of the model error measurements, etc.), for the purposes of demonstrating the expandability of the data diff if two models pass the ANCOVA tests, they will be considered equivalent. It is also important to note that the comparison is implemented as a workflow within e-Science Central and, as such, can be expanded to include additional tests if necessary. 

\end{description}

\begin{figure}[htbp]
   \centering
   \includegraphics[scale=0.15]{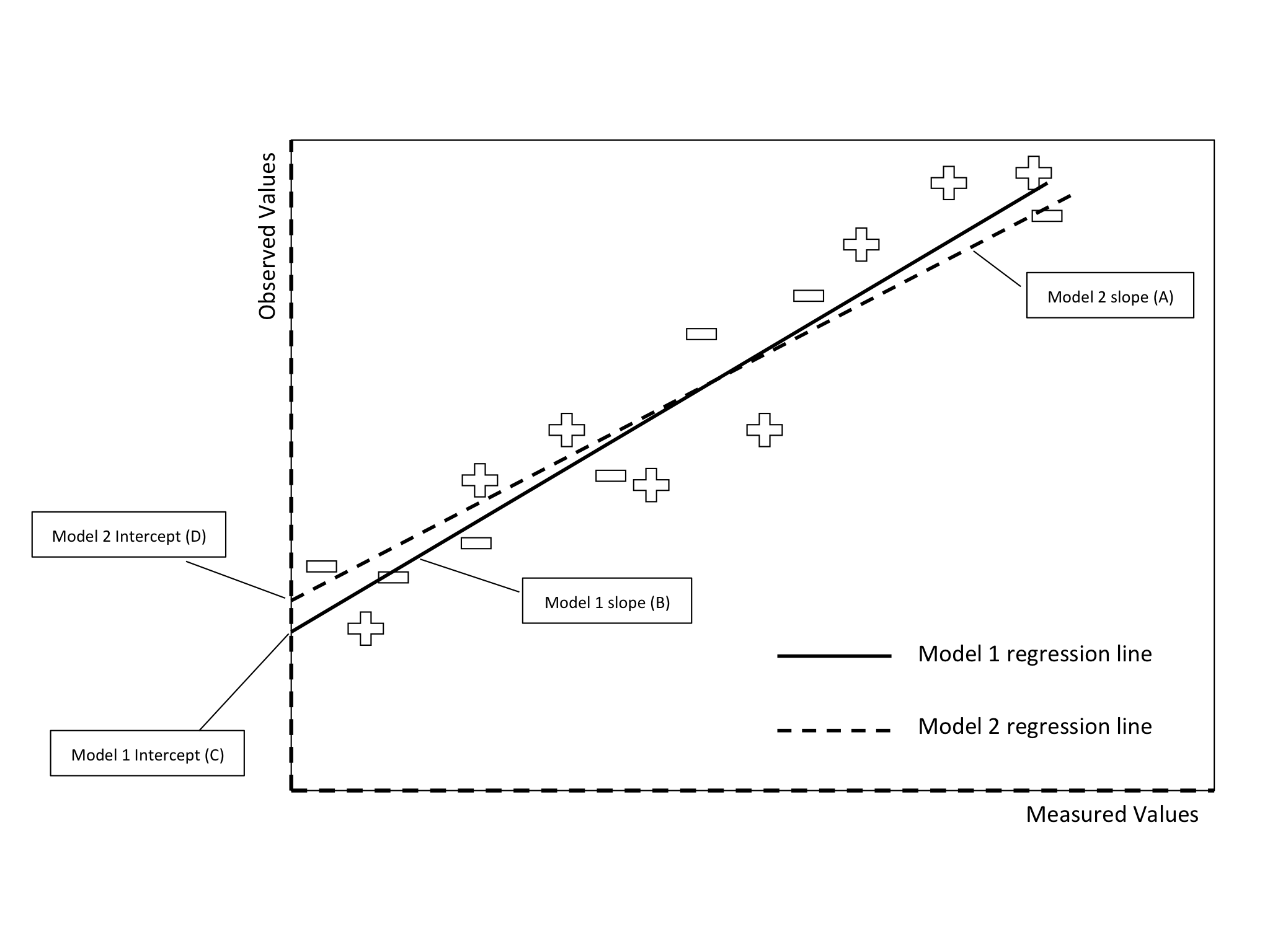} 
   \caption{Comparison of Model fit}
   \label{fig:modelfit}
\end{figure}

\section{Infrastructure for reproducibility in e-Science Central}  \label{sec:infrastructure}

\begin{figure}[htbp]
   \centering
   \includegraphics[scale=0.6]{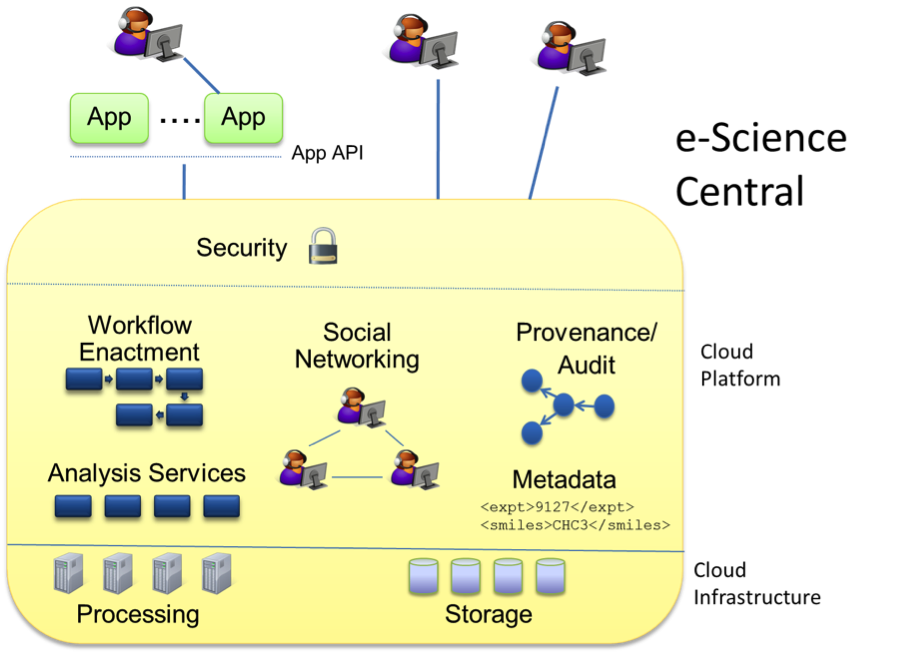} 
   \caption{e-Science Central Components}
   \label{fig:esc_arch}
\end{figure}

The \esc system~\cite{pw2009} is a portable cloud `platform-as-a-service' that can be deployed on either private clusters or public clouds, including Amazon EC2 and Windows Azure. Cloud computing has the potential to give scientists the computational resources they need, when they need them. However, cloud computing does not make it easier to build the often complex, scalable secure applications needed to support science. e-Science Central was designed to overcome these obstacles by providing a platform on which users can carry out their research, and build high-level applications. Figure \ref{fig:esc_arch} shows the key components of the e-Science Central `Platform as a Service' sitting on an `Infrastructure as a Service' cloud. It combines three technologies -- Software as a Service (so users only need a web browser to do their research), Social Networking (to support sharing and community interaction) and Cloud Computing (to provide storage and computational power). Using only a browser, users can upload data, share it in a controlled way with colleagues, and analyse the data using either a set of pre-defined services, or their own, which they can upload for execution and sharing. A range of data analysis and programming  languages is supported, including Java, Javascript, R and Octave. From the point of view of users, this gives them the power of cloud computing without them actually having to manage the complexity of developing cloud-specific software -- they can create services in a variety of languages, upload them into e-Science Central, and have them run transparently on the cloud.  These services can be composed into workflows to automate the analysis.

\begin{figure*}[htbp]
   \begin{center}
      \includegraphics[scale=0.7]{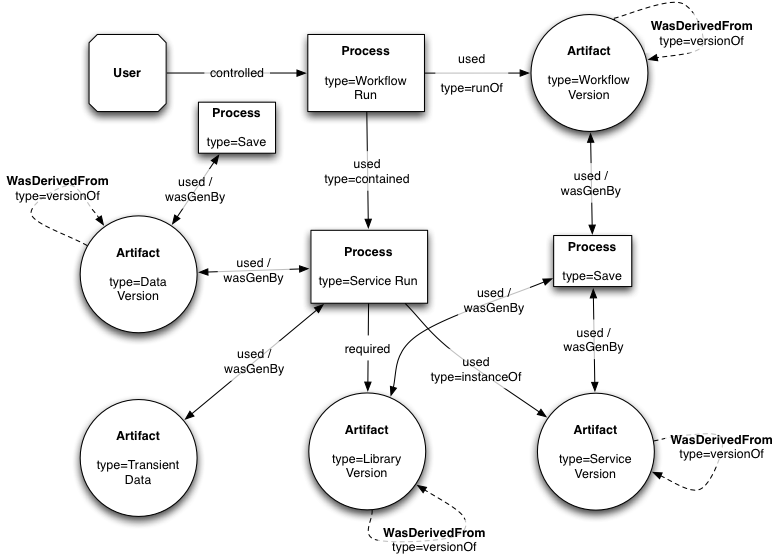} 
   \caption{\esc Provenance Model}
   \label{fig:prov_model}
   \end{center}
\end{figure*}

Versioning is an integral storage feature in e-Science Central, allowing users to work with old versions of data, services and workflows. All objects (data sets, services and workflows) are stored in files through a virtual filestore driver than can be mapped onto a range of actual storage systems including standard local and distributed filesystems, and Amazon S3. When a file is stored, if a previous version exists then a new one is automatically created.  

Whilst three of the elements, workflows, services and libraries (the latter two being part of $ED$), presented in Section \ref{sec:framework} are within the control of \esc and are versioned this alone is not sufficient to detect workflow decay.  By default, workflow $W$ will use the latest version of a service, $S_{nVi}$ when it is executed at time $t$.  However at time, $t'$ a service version may have been updated and so $S_{nVj}$ where $j > i$ would be used.  It is possible to override this behaviour and always use a particular version of a service if desired.  In addition, even if a workflow evolves $W_i \rightarrow W_j$ the evolution may only effect branches which do not appear in a particular execution trace.

\subsection{Capturing provenance traces in e-Science Central}\label{sec:capturing}

The provenance system within e-Science Central is used to capture the history and life cycle of every piece of data within the system.  For instance: who created the data? who has downloaded the data? what version of which services (in what workflow) accessed the data? and who has the data been shared with?

The provenance data model, shown in Figure \ref{fig:prov_model}, is based on the Open Provenance Model (OPM) Version 1.1~\cite{Moreau2010a}, and can be used to produce a directed acyclic graph of the history of an object.  Objects in OPM are categorised as either artifacts, processes or actors which correspond to nodes within the graph.  Vertices in the graph represent relationships between two objects and are of types such as wasGeneratedBy, used, wasControlledBy and wasDerivedFrom. The OPM Core model has been extended with subclasses to identify the different types of processes and artifacts we are concerned with. For example, execution of a workflow and a service within a workflow have been differentiated.  The relationship between workflow execution and service execution is of type \emph{contained} and is not strictly required by our model but its inclusion makes it easier to generate the different views of the process. The artifacts described in the model are also subclassed in order to differentiate between versions of data, services, libraries and workflows.  Without this subclassing, it would be necessary to either encode the object type in the identifier (not desirable as it adds an unnecessary layer of obfuscated information) or perform many lookups to answer the question `what type of object has a particular identifier'.  

The relationships between  processes and artifacts is specified in the OPM standard with the \emph{Save} process taking one version of an artifact and generating a new version.  

The self referential relationship between artifacts of type \emph{WasDerivedFrom}, shown with a dashed line, indicates a second level, inferred relationship omitting the \emph{Save} process.   This implies that \Div is in some way derived from D$_{i,v-1}$ when $v > 0$.  

Our model deals with two different types of data artifact: \emph{Data Version} and \emph{Transient Data}.  This is due to the semantics of  workflows in \esc: data generated by a workflow must be explicitly saved using a service which `exports' the data back into the \esc repository.  Any data which is not explicitly exported will be discarded when the workflow completes.  

\subsection{Provenance database implementation} \label{sec:implementation}

Given that the provenance structure being stored is a directed acyclic graph, we chose to store it in the non-relational graph database, Neo4j (\url{www.neo4j.org}). Neo4j differs from traditional relational databases as its structure is not in terms of tables and rows and columns, but in nodes, relationships and properties.  This provides a much more natural fit to our model, and allows us to store the provenance graph directly instead of encoding it in a relational model.  Neo4j has also been shown to scale well, and most importantly, to perform well in terms of queries even when storing a very large graph structure~\cite{Dominguez-Sal:2010:SGD:1927585.1927590}.  Libraries are provided to allow users to work with Neo4j directly in either Java or Ruby.  

Various components within \esc all log provenance information.  From the outset the system was designed such that the order in which provenance events are received by the server is not important. For instance, we could have an arbitrary interleaving of the events which signify that a `Service has run' and that the `Service accessed some data'.  Each of these events contains a subset of information which must be added to the provenance database:  the first event contains the service name, start and completion time (and some other information) whereas the latter defines the identity of the data which was read.  Instead of SQL queries, Neo4j supports an operation known as a \emph{traversal} whereby the user defines a starting node and rules about what types of relationship/node to `traverse'.  The result is a set of paths from the starting node to the acceptable ending nodes.  The provenance server is decoupled from \esc with a durable JMS queue which allows \esc to minimise the number of synchronous write operations which must be performed in a synchronous request.  

\subsection{Implementing PDIFF using Neo4j}

The Neo4j database used by \esc contains not just the execution traces which we are interested in traversing for PDIFF but also all the provenance for other items within \esc.  For example, there may be multiple executions of each workflow but we are only interested in two specific executions.  Also, the provenance may indicate a data set may have been created through a \emph{Save} process which does not concern us.

The Neo4j traversal framework allows us to traverse the graph but restrict the traversal to relationships of a particular \emph{type} and \emph{direction}.  During most of our traversals we restrict ourselves to the $\USED$ and $\GENBY$ types with direction $\mathit{outgoing}$.  This allows us to navigate back from the output data set in each execution trace back to the original input data sets.  However, the semantics of Neo4j traversals are that you are guaranteed to visit each node in the graph but not traverse every edge.  This would have the potential to miss branches in the execution trace which we should be comparing.  To mitigate this we make additional calls to the database to ensure we retrieve all the relationships for each node.

As we traverse through the graph we build up a report containing the tree of `Delta' descriptions making use of the functions described below.

\begin{description}

\item [\textproc{upD}($d$)] Navigates from either a Transient Data or Data Version node along an outgoing $\GENBY$ relationship to the Service Run node which created it and returns this node.  If there is no such relationship null is returned indicating the end of the graph.  In this case, $d$ is a Transient Data node and an exception is thrown as this is an invalid starting point in the execution trace.  

\item [\textproc{upS}($a,p$)] Navigates along an outgoing $\USED$ relationship to a Transient Data or Data Version node and returns it.  The Neo4j relationship which represents the edge contains an attribute for the port name $p$ and so we filter based on this.  If no such node is found an exception is thrown as this is an invalid starting point to the execution trace.

\item [\textproc{dDiff(nL, nR)}] If $nL$ and $nR$ are Transient Data nodes then this returns whether or not the MD-5 hash values of the data are the same.  If $nL$ and $nR$ are Data Version nodes then it will execute and \esc workflow to determine whether or not the data sets are equivalent.  The workflow executed is dependent on the MIME type of the data set as recorded as an attribute of the node in Neo4j.

\item [\textproc{aDiff(nL, nR)}]  This returns a 2-tuple containing whether or not the service identifier and version identifier for $nL$ and $nR$ are equal.  In order to get the service and version identifier we must navigate the $\mathit{instanceOf}$ relationship to the Service Version node in the provenance database.  If either of the Service Version nodes is null then an exception is thrown as the graph is invalid.

\item [\textproc{findNode(nL, nR)}]  This performs a breadth first search of each of the execution traces in order to try to find two nodes that enable the analysis to be re-synchronised.  For brevity we will assume in the following discussion that the Service Run nodes can be compared directly -- in practice we must navigate along the $\mathit{instanceOf}$ relationship and perform the comparison there.  Initially we select $nL'$ as a match candidate which is the next Service Run in the trace traversing along $\USED$ and $\GENBY$ with direction $\mathit{outgoing}$.  We then perform a breadth first traversal of the same relationship types beginning from $nR$ searching for a match with $nL'$.  If a match is found it is recorded as $candidateL$ along with the depth at which it was found.  If no match was found, we traverse back one step further from $nL'$ and repeat the search from $nR$.

If no candidate is found, the entire graph fragments from $nL$ and $nR$ are returned as there is no matching node that enables the graphs to be re-synchronised.  However, if a match is found, the search is re-performed but searching for $nR'$ nodes in the $nL$ graph.  When a second match is found, namely $candidateR$,  the search terminates.  The returned nodes (and associated graph fragments) are those with the lowest combined depths from $nL$ and $nR$.  The second search deals with the case where a closer node match is found had we searched the other graph first.  We can terminate on finding a match as we know that this will either be the same match as the $candidateL$ or a match with a lower combined depth.  The nodes which are returned as a synchronisation point are also stored so that we are able to join future branches in the results. 

\item [\textproc{addDelta($\Delta$, dataElement)}]  Builds up the result graph.  The results are constructed using a simple graph implementation where each node is able to contain one of the following objects: a Transient Data mismatch (the actual MD-5 hash values are of little use); the service and version identifiers in case of a service or version change; or two graphs representing the graph fragments which appear in one of the graphs but not the other.  In the case of the graph fragments, the previous node $nL$ and $nR$ are included along with the subsequent nodes $nL'$ and $nR'$ for clarity.  The result is stored in \esc and can be exported to GraphML for visualisation or use in other applications.

\end{description}

At this time, the \PDIFF implementation is not yet integrated into \esc. In the next stage, the diagnosis functionality offered  by the algorithm will be exposed to users through a simple interface. Users will be able to select two workflow runs and obtain a ``diff'' report on their divergence. Thanks to the relative small size of typical workflows, the diff graph can be visualized as part of the \esc web-based graphical interface.

\section{Conclusions}  \label{sec:conclusion}

The ability to reproduce experimental results underpins scientific discourse and is key to establishing credibility in scholarly communication. Yet, several factors complicate the task of reproducing other scientists' results in settings other than the original one. Such factors include differences in the experimental environment (different lab, research group, equipment technology) and insufficient detail in the description of the methods. 
In computational science, the operating environment typically includes e-science infrastructure facilities, and the methods are formally encoded in the form of executable scripts, workflows, or general computer programs. This level of automation creates the perception that the limitations in reproducibility of results can be easily overcome. In reality, however, the traditional limiting factors are simply replaced by others, including an unstable and evolving operational environment, and lack of maintenance of related programs with mutual dependencies, which become unable to function together. As a result, scientific applications suffer from forms of decay that limit their longevity, and thus their reuse and evolution in time.

In this paper we have focused specifically on workflow-based e-science, and on a scenario where attempts to reproduce earlier results translates into new runs of the same workflow at a later time. We assume that the workflow is still executable, but it may produce unexpected results or may have been made dysfunctional by changes in its system environment, or by uncontrolled evolution of the workflow itself or of its input data.

We have presented two main contributions. Firstly, we have proposed a framework meant to clarify the range of meanings of the often ambiguous term ``reproducibility". We have then used the famework to situate our second contribution, namely the \PDIFF~algorithm for diagnosing the divergence between two executions by comparing their associated provenance traces. We have also described a number of data comparison function, which underpin provenance comparison. Finally, we have discussed the current implementation of \PDIFF~on the \esc workflow management middleware. \PDIFF~will be integrated in the public-facing version of \esc in the near future, with attention to usability testing, which has not being addressed in the current prototype version.

\bibliographystyle{wileyj}
\bibliography{reproducibility-CCPE-2012,sjw}

\end{document}